\documentclass{article}%
\usepackage{amsmath}
\usepackage{amsfonts}
\usepackage{amssymb}
\usepackage{graphicx}%
\providecommand{\U}[1]{\protect\rule{.1in}{.1in}}

\begin{document}

\begin{center}
\textbf{BRANS-DICKE WORMHOLE REVISITED}

\bigskip

\bigskip

\textbf{Amrita Bhattacharya}$^{1,a}$\textbf{, Ilnur Nigmatzyanov}$^{2,b}%
$\textbf{, Ramil Izmailov}$^{2,c}$

\textbf{and}

\textbf{Kamal K. Nandi}$^{1,2,3,d}$

$\bigskip$

$\bigskip$

$^{1}$Department of Mathematics, University of North Bengal, Siliguri (WB)
734013, India

$^{2}$Joint Research Laboratory, Bashkir State Pedagogical University, Ufa
450000, Russia

$^{3}$Department of Theoretical Physics, Sterlitamak State Pedagogical
Academy, Sterlitamak 453103, Russia

$\bigskip$

$\bigskip$

$\bigskip$

$^{a}$Email: amrita\_852003@yahoo.co.in

$^{b}$Email: ilnur.nigma@gmail.com

$^{c}$Email: izmailov.ramil@gmail.com

$^{d}$Email: kamalnandi1952@yahoo.co.in

\bigskip

\textbf{Abstract}
\end{center}

A basic constraint to be satisfied by Brans class I solution for being a
traversible wormhole is derived. It is argued that the solution could be a
\textit{wormhole analogue} of the Horowitz-Ross naked black hole. It is
further demonstrated that the wormhole is traversible only \textquotedblleft
in principle\textquotedblright, but not in practice. Using a recently proposed
measure of total gravitational energy inside a static wormhole configuration,
it is shown that the wormhole contains repulsive gravity required for the
defocussing of orbits at the throat.

\bigskip

PACS number(s): 04.20.Gz, 04.62.+v

\bigskip


\bigskip

\textbf{I. Introduction}

Lorentzian wormholes are objects of research in the frontier of theoretical
physics. Einstein's gravitational field equations predict not only black
holes, but also wormholes and neither of these has yet been ruled out by
observations. Wormhole physics has lately gained a renewed impetus following
the pioneering work by Morris, Thorne and Yurtsever [1] in 1988, although
wormhole solutions were conceived as particle models by Einstein himself
(Einstein-Rosen bridge [2]) in 1935. Wheeler's geometrodynamics [3] tells us
that wormholes are like topological handles connecting two distant regions of
spacetime (like Klein bottle) or even two distinct universes. In Ref[1], a
general theoretical framework for these handles have been formulated and
conceived as possible tunnels facilitating rapid interstellar travel. Over the
years, however, newer possibilities have emerged. For instance, these objects
offer an intriguing possibility as to whether they might act as effective
gravitational lenses in astrophysical scenarios. This has been conjectured by
Cramer \textit{et al}. [4], who recommended an analysis of massive compact
halo objects (MACHOs) search data for the detection of such lens effects. Such
data also offer a distinct possibility which would allow us in future to
distinguish between the strong field lensing effects caused by macroscopic
wormholes and black holes [5-8]. Could it be that the early universe was
populated by microsopic semiclassical wormholes? Though the answer is not
definitively known, an analysis of the consequences can be found in Ref.[9].
Appearance of microscopic wormholes were speculated even in the popular
articles on recent (unfinished) high energy LHC experiment. Kardashev, Novikov
and Shatskiy [10] suggest that black holes could be past wormhole entrances.
Undeniably, all these effects are still at a very speculative level, yet they
are quite promising ones. The unique treatise by Visser [11] provides a
detailed account of classical and quantum gravity wormholes.

An arena for the \textit{natural} occurrence of classical Lorentzian wormholes
is the Brans-Dicke (BD) theory which involves a scalar field. It might be
recalled here that, in general, scalar fields have come to stay with us not
only because of the Machian nature of BD theory, but also because the string
theory, higher dimensional theories or $f(\mathbf{R})$ gravity theories
invariably predict their own modulo scalar fields. In the BD theory, a search
for static wormholes has been initiated by Agnese and La Camera [12] who have
shown that the BD scalar can play the role of exotic matter provided the
coupling parameter $\omega<-2$. This work has been followed by investigations
in other classes of Brans solutions [13] as well as in the Einstein
conformally rescaled BD theory [14]. Several related works also exist in the
literature, see for instance [15-23] $-$ to mention only a few. Considering
the importance of BD theory in the interpretation of various astrophysical
phenomena, it is important that a thorough analysis of static spherically
symmetric wormhole solutions be undertaken in this theory. This provides the
motivation for the present paper.

The purpose of the present article is to report the following investigations:
We derive the basic constraint leading to the ranges for $\omega$ for which
traversible wormhole in the BD theory could be possible. One of the two ranges
turn out to be different from that derived by Agnese and La Camera [12]. We
discuss the violation of energy conditions on the basis of these ranges. A
particularly interesting result is that the Horowitz-Ross naked black hole,
defined later, has a wormhole analogue in the Brans class I solution. The
impact of the ranges of $\omega$ on traversibility is explicitly demonstrated.
A recent definition of gravitational energy is also implemented in the BD
wormhole to show that it contains repulsive gravity, as required.

The paper is organized as follows: In Sec.II, we develop the usual geometry of
Morris-Thorne wormholes for the Brans I solution. In Sec.III, we derive ranges
of $\omega$ and discuss the violation of energy conditions. In Sec.IV, we
discuss traversibility and in Sec.V, we compute the gravitational energy
present in the wormhole. Sec.VI presents the conclusions and some relevant remarks.

\textbf{II. Brans I wormhole solution}

The matter-free action in the Jordan frame is ($8\pi G=c=1$, unless
specifically restored)%
\begin{equation}
S=\frac{1}{2}\int d^{4}x(-g)^{\frac{1}{2}}\left[  \varphi\mathbf{R}%
-\varphi^{-1}\omega(\varphi)g^{\mu\nu}\varphi_{,\mu}\varphi_{,\nu}\right]  .
\end{equation}
The field equations are%
\begin{equation}
\square^{2}\varphi=0,
\end{equation}

\begin{equation}
\mathbf{R}_{\mu\nu}-\frac{1}{2}g_{\mu\nu}\mathbf{R}=-\frac{\omega}{\varphi
^{2}}\left[  \varphi_{,\mu}\varphi_{,\nu}-\frac{1}{2}g_{\mu\nu}\varphi
_{,\sigma}\varphi^{,\sigma}\right]  -\frac{1}{\varphi}\left[  \varphi_{;\mu
}\varphi_{;\nu}-g_{\mu\nu}\square^{2}\varphi\right]  ,
\end{equation}
where $\omega$ is a constant dimensionless coupling parameter, $\square
^{2}\equiv(\varphi^{;\rho})_{;\rho}$ and $\varphi$ is the Brans-Dicke scalar.
The general solution in isotropic coordinates ($t,r,\theta,\varphi$) is given
by%
\begin{equation}
d\tau^{2}=-e^{2\alpha(r)}dt^{2}+e^{2\beta(r)}dr^{2}+e^{2\nu(r)}r^{2}%
(d\theta^{2}+\sin^{2}\theta d\psi^{2}).
\end{equation}
Brans derived four classes of solutions (I-IV), but the most widely used one
is the class I solution [24]. It corresponds to a gauge $\beta-\nu=0$ and is
given by%
\begin{equation}
e^{\alpha(r)}=e^{\alpha_{0}}\left[  \frac{1-B/r}{1+B/r}\right]  ^{\frac
{1}{\lambda}},
\end{equation}%
\begin{equation}
e^{\beta(r)}=e^{\beta_{0}}\left[  1+B/r\right]  ^{2}\left[  \frac
{1-B/r}{1+B/r}\right]  ^{\frac{\lambda-C-1}{\lambda}},
\end{equation}%
\begin{equation}
\varphi(r)=\varphi_{0}\left[  \frac{1-B/r}{1+B/r}\right]  ^{\frac{C}{\lambda}%
},
\end{equation}%
\begin{equation}
\lambda^{2}=(C+1)^{2}-C\left(  1-\frac{\omega C}{2}\right)  >0,
\end{equation}
where $\alpha_{0}$, $\beta_{0}$, $B$, $C$, and $\varphi_{0}$ are constants.
The constants $\alpha_{0}$ and $\beta_{0}$ are determined by asymptotic
flatness condition as $\alpha_{0}=$ $\beta_{0}=0$.

In order to investigate the possibility if the above solution represents
wormholes, it is convenient to cast the spacetime geometry in the
Morris-Thorne canonical coordinates%
\begin{equation}
d\tau^{2}=-e^{2\Phi(R)}dt^{2}+\left[  1-\frac{b(R)}{R}\right]  ^{-1}%
dR^{2}+R^{2}(d\theta^{2}+\sin^{2}\theta d\psi^{2})
\end{equation}
where $\Phi(R)$ and $b(R)$ are redshift and shape functions respectively.
These functions are required to satisfy some constraints, enumerated in [1],
in order that they represent a wormhole. It is, however, important to stress
that the choice of coordinates is purely a matter of convenience and not a
physical necessity. For instance, one could equally well work directly with
isotropic coordinates or proper quantities using the expressions in Ref.[11],
but the final conclusions would be the same.

Redefining the radial coordinate $r\rightarrow R$ in the metric (4) as%
\begin{equation}
R=re^{\beta_{0}}\left[  1+B/r\right]  ^{2}\left[  \frac{1-B/r}{1+B/r}\right]
^{\Omega}\text{, \ \ \ \ \ \ \ \ \ }\Omega=1-\frac{C+1}{\lambda},
\end{equation}
we obtain the following functions for $\Phi(R)$ and $b(R)$:%
\begin{equation}
\Phi(R)=\alpha_{0}+\frac{1}{\lambda}\left[  \ln\left\{  1-\frac{B}%
{r(R)}\right\}  -\ln\left\{  1+\frac{B}{r(R)}\right\}  \right]  ,
\end{equation}%
\begin{equation}
b(R)=R\left[  1-\left\{  \frac{\lambda\{r^{2}(R)+B^{2}\}-2r(R)B(C+1)}%
{\lambda\{r^{2}(R)-B^{2}\}}\right\}  ^{2}\right]  .
\end{equation}
The throat of the wormhole occurs at $R=R_{0}$ such that $b(R_{0})=R_{0}$.
This gives the minimum allowed $r$-coordinate radii $r_{0}^{\pm}$ as%
\begin{equation}
r_{0}^{\pm}=\alpha^{\pm}B,
\end{equation}%
\begin{equation}
\alpha^{\pm}=(1-\Omega)\pm\sqrt{\Omega(\Omega-2}.
\end{equation}
The values $R_{0}^{\pm}$ can be obtained from Eq.(10) using this $r_{0}^{\pm}%
$. Noting that $R\rightarrow\infty$ as $r\rightarrow\infty$, we find that
$b(R)/R\rightarrow0$ as $R\rightarrow\infty$. Also, $b(R)/R\leq1$ for all
$R\geq$ $R_{0}^{\pm}$. The redshift function $\Phi(R)$ has a singularity at
$r=r_{S}=B$. Therefore, in order that the wormhole be just geometrically
traversible, the minimum allowed values $r_{0}^{\pm}$ must exceed $r_{S}=B$.
We shall see below under what values of the coupling constant $\omega$ it is possible.

\textbf{III. Violation of energy conditions}

The expressions for the stress components in the orthonormal rest frame of the
observer are [1]%
\begin{align}
\rho &  =\frac{b^{\prime}}{R^{2}}\\
p_{R}  &  =2\left(  1-\frac{b}{R}\right)  \frac{\Phi^{\prime}}{R}-\frac
{b}{R^{3}}\\
p_{\theta}  &  =p_{\psi}=\left(  1-\frac{b}{R}\right)  \left[  \Phi
^{\prime\prime}+\Phi^{\prime2}+\frac{\Phi^{\prime}}{R}-\frac{b^{\prime}%
R-b}{2R^{2}}\left(  \Phi^{\prime}+\frac{1}{R}\right)  \right]  ,
\end{align}
where primes denote differentiation with respect to $R$. Thus the energy
density of the wormhole material is given by%
\begin{equation}
\rho(R)=(R^{-2})(db/dR)
\end{equation}
so that
\begin{equation}
\rho=-\frac{4B^{2}r^{4}Z^{2}[(C+1)^{2}-\lambda^{2}]}{\lambda^{2}(r^{2}%
-B^{2})^{4}},
\end{equation}
where
\begin{equation}
Z\equiv\left(  \frac{r-B}{r+B}\right)  ^{(C+1)/\lambda}.
\end{equation}

To have a wormhole spacetime, we shall allow the source scalar field to
violate one or more energy conditions, particularly the Weak Energy Condition
(WEC) $\rho>0$ and/or the Null Energy Condition (NEC) $\rho+p_{R}\geq0$ where
$\rho$ is the matter energy density and $p_{R}$ is the radial pressure.
(Transverse pressures $p_{\theta}$, $p_{\psi}$ are not considered as they
refer strictly to ordinary matter.) The violation of NEC is a minimal
requirement to have defocussing of light trajectories (\textit{repulsive}
gravity) passing across the wormhole throat. Since it is a minimal
requirement, a stronger violation (namely, of WEC) allowing defocussing is not
ruled out. The necessity of NEC violation in wormholes is provided by the
Topological Censorship Theorem [25] and by dynamical circumstances [26].

At $r=$ $r_{S}=B$, the density $\rho$ diverges because it is a surface where
all curvature invariants diverge (naked singularity). However, this surface is
inaccessible to a traveller as the minimum radius he/she reaches during travel
is $r_{0}^{\pm}>r_{S}$. We see from Eq.(19) that the WEC is violated at all
$r$ including the throat under the condition
\begin{equation}
(C+1)^{2}>\lambda^{2}.
\end{equation}

Next we want to derive the specific dependence of the constraints on the
coupling parameter $\omega$. For this purpose, henceforth we shall consider
the weak field value of $C$, viz.,%
\begin{equation}
C=-\frac{1}{\omega+2}.
\end{equation}
Putting this in Eq.(8), we get%
\begin{equation}
\lambda=\pm\sqrt{\frac{2\omega+3}{2\omega+4}}.
\end{equation}

(i) Let us consider the $+$ sign before $\lambda$. In the limiting case,
$C(\omega)\rightarrow0$, $\lambda(\omega)\rightarrow1$ as $\omega
\rightarrow\pm\infty$, one simply recovers the Schwarzschild exterior metric
in standard coordinates from Eqs.(9),(11) and (12), so that $b(R)=2M$ and
$\rho=0$ (Schwarzschild exterior vacuum). It is clear that $\lambda^{2}>0$ for
all $\omega$ except in the range $-2<\omega<-\frac{3}{2}$, and hence we
exclude it from consideration. We find that the wormhole condition, viz.,
$(C+1)^{2}>\lambda^{2}$ is satisfied only in the range $\omega<-2$. This
implies $\gamma=C+1=\frac{\omega+1}{\omega+2}>1$, which is exactly the result
obtained by Agnese and La Camera [12]. For $\omega<-2$, it can also be checked
from Eq.(13) that, out of the two throat radii $r_{0}^{\pm}$, only $r_{0}%
^{+}>B$, as is required to avoid the singular surface. For $\omega>-2$, which
we have already excluded, the throat radii become negative.

To analyze the energy conditions at and away from the throat, let us compute
the radial pressure, which, using Eqs.(11), (12) in (16), turn out to be%
\begin{equation}
p_{R}=-\frac{4Br^{3}Z^{2}}{\lambda^{2}(r^{2}-B^{2})^{4}}[\lambda C(r^{2}%
+B^{2})-Br(C^{2}-1+\lambda^{2})].
\end{equation}
Adding Eqs.(19) and (24), we get%
\begin{equation}
\rho+p_{R}=-\frac{4Br^{3}Z^{2}}{\lambda^{2}(r^{2}-B^{2})^{4}}[\lambda
C(r^{2}+B^{2})+2Br(C+1-\lambda^{2})].
\end{equation}

Let us concentrate on Eq.(13) and the inequality (21). We find from Fig.1 that
$r_{0}^{+}>B$ and $f=(C+1)^{2}-\lambda^{2}>0$ hold simultaneously when
$\omega<-2$ . Also Fig.2 shows that $\rho<0$, $\rho+p_{R}<0$ for a typical
value, say, $\omega=-5$ and for all values of $r\geq r_{0}^{+}$. These imply
that both WEC and NEC\ are violated all the way up from the throat. Moreover,
$\rho\rightarrow0$, $\rho+p_{R}\rightarrow0$ as $r\rightarrow\infty$ (which is
the same as $R\rightarrow\infty$), in accordance with the asymptotic flatness
of the BD solution under consideration.

(ii) Next let us consider the $-$ sign before $\lambda$. In the limiting case,
$C(\omega)\rightarrow0$, $\lambda(\omega)\rightarrow-1$ as $\omega
\rightarrow\pm\infty$. This has the effect that we can no longer recover the
Schwarzschild solution because $g_{00}$ and $g_{11}$ diverge at $r=B$.
However, we shall concentrate only on the finite values of $\omega$ allowing
wormhole spacetime. It turns out that within a small range, $-\frac{3}%
{2}<\omega<-\frac{4}{3}$, wormholes \textit{are} possible. From Fig.3, it
follows that the conditions $r_{0}^{+}>B$ and $f=(C+1)^{2}-\lambda^{2}>0$ are
simultaneously satisfied in that range. Also Fig.4 shows that $\rho<0$,
$\rho+p_{R}<0$ for a typical value, say $\omega=-1.4$ and for all $r\geq
r_{0}^{+}$ . In other words, all the features of case (i) exist in this case too.

The above analysis indicates that $(C+1)^{2}>\lambda^{2}$ is a \textit{basic}
constraint to be satisfied by the BD constants, supporting the discussion in
Refs.[13,14,27,28]. This concludes our analysis about the allowed values for
$\omega$, that is, depending on the sign of $\lambda$, wormholes are possible
when $\omega<-2$ as in (i) and $-\frac{3}{2}<\omega<-\frac{4}{3}$ as in (ii).

\textbf{IV. Traversibility}

Does the BD solution under consideration represent a humanly traversible
wormhole? In general, there are several constraints for human traversibility
as discussed in Ref.[1]. One condition for traversibility is that the
spacetime be asymptotically flat, which is indeed the case with the solution
(4). However, the most important condition is that the tidal forces be finite
and tolerable everywhere, especially at the throat.

The differential of the radial tidal acceleration $\Delta a^{r}$ in the static
orthonormal frame ($\widehat{e}_{t,}\widehat{e}_{R,}\widehat{e}_{\theta
,}\widehat{e}_{\varphi}$) is given by%
\begin{equation}
\Delta a^{r}=-\mathbf{R}_{\widehat{R}\widehat{t}\widehat{R}\widehat{t}}\xi
^{R},
\end{equation}
where $\xi^{R}$ is the radial component of the separation vector and the
curvature tensor component is given by [1]%
\begin{equation}
\left\vert \mathbf{R}_{\widehat{R}\widehat{t}\widehat{R}\widehat{t}%
}\right\vert =\left\vert (1-b/R)\left\{  -\Phi^{\prime\prime}+\frac{b^{\prime
}R-b}{2R(R-b)}\Phi^{\prime}-(\Phi^{\prime})^{2}\right\}  \right\vert .
\end{equation}
This component is invariant under a Lorentz boost [1,29]. For the metric given
by Eqs.(11) and (12), we find in the freely falling orthonormal frame
($\widehat{e}_{0^{\prime},}\widehat{e}_{1^{\prime},}\widehat{e}_{2^{\prime}%
,}\widehat{e}_{3^{\prime}}$) with velocity $v$ the following expression%
\begin{equation}
\left\vert \mathbf{R}_{\widehat{1}^{\prime}\widehat{0}^{\prime}\widehat
{1}^{\prime}\widehat{0}^{\prime}}\right\vert =\left\vert \mathbf{R}%
_{\widehat{R}\widehat{t}\widehat{R}\widehat{t}}\right\vert =\left\vert
\frac{4Br^{3}Z^{2}[\lambda(r^{2}+B^{2})-Br(C+2)]}{\lambda^{2}(r^{2}-B^{2}%
)^{4}}\right\vert .
\end{equation}

A point must be clarified here. In Ref.[11], the components of the Riemann
tensor are given in mixed (1,3) form whereas we are using here the purely
covariant (0,4) form, following the notation in Refs.[1,29]. This does not
make any difference; we could as well use the mixed notation. Actually, the
expression in Eq.(27) as well as all other components are exactly the same as
the ones given in Ref.[11]. This happens because both are calculated in the
orthonormal frame in which raising and lowering of indices is done only by the
locally flat Minkowski metric $\eta_{\mu\nu}=(-1.+1,+1,+1)$.

We now put the values of $C(\omega)$ and $\lambda(\omega)$ from Eqs (22), (23)
in the right hand side of Eq.(28) and write $\left\vert \mathbf{R}%
_{\widehat{1}^{\prime}\widehat{0}^{\prime}\widehat{1}^{\prime}\widehat
{0}^{\prime}}\right\vert \equiv g(\omega,r,B)$. Let us vary $r$ away from the
throat for fixed values of $\omega$ within the ranges derived in the previous
section. To be suitable for comfortable travel by a human of length $2$
$mtrs$, the tidal acceleration should be roughly of the order of one Earth
gravity $g_{\oplus}$ ($=\frac{GM_{\oplus}}{r_{\oplus}^{2}}\simeq980$
cm/sec$^{2}$). In relativistic units, it should be $\frac{g_{\oplus}}{c^{2}}$.
Since $\xi^{R}$ has the dimension $L$ of length, looking at Eq.(26), we get
that the dimension of $\mathbf{R}_{\widehat{R}\widehat{t}\widehat{R}%
\widehat{t}}$ must be $\frac{g_{\oplus}}{c^{2}\times L}\sim L^{-2}$.
Therefore, the right hand side Eq.(28) should be less than $\lesssim
\frac{g_{\oplus}}{c^{2}\times2mtrs}\sim10^{-20}cm^{-2}$ [1], which requires
that the magnitude of the curvature components should be very close to zero in
the orthonormal frame of the traveller. Such a condition is easily provided by
$\Phi=0$, which is not the case here. To have an idea of the magnitudes
involved, we focus on a typical value in case (i), say, $\omega=-5$ and
observe the following:

The plot of $g(r)$ vs $r$ shown in Fig.5 in units $B=1$ reveals a remarkable
feature of Brans class I wormhole. At the throat location, $r_{0}^{+}=1.958$,
we see that $g(r_{0}^{+})$ is of the order of $10^{-3}cm^{-2}$, which
\textit{increases} up to $g(r)=10^{-2}cm^{-2}$ at a location $r=2.313$ away
from the throat! Thereafter, $g(r)$ decays rapidly. That is, an infalling
observer meets the maximum radial tidal force not at the throat but above it.
This phenomenon is very much analogous to the idea of naked black holes first
discussed by Horowitz and Ross [29]. They defined the naked black hole as a
spacetime in which an infalling observer meets the maximum tidal force not at
the horizon but above it. In a freely falling frame, the curvature components
could be larger than those at the horizon. Since the region of large tidal
forces is visible to distant observers, Horowitz and Ross called such objects
\textquotedblleft naked black holes.\textquotedblright\ In our case, the role
of horizon is played by the throat of the analogue wormhole. However, no such
phenomenon occurs for the range in case (ii) $-$ there is a steady decrease in
curvature right from the throat up, as can be seen from Fig.6.

As we read off from Fig.5, the maximum tidal force unfortunately is still
$10^{18}$ times more than that on Earth's gravity field. Hence the wormhole is
not humanly traversible. Since human travel is out of question, we might try
to calculate the speed $v$ of an inanimate test particle passing through the
wormhole. The lateral tidal forces in the Lorentz boosted frame of the
particle are [29]%
\begin{align}
\mathbf{R}_{\widehat{2}^{\prime}\widehat{0}^{\prime}\widehat{2}^{\prime
}\widehat{0}^{\prime}}  &  =\mathbf{R}_{\widehat{\theta}\widehat{t}%
\widehat{\theta}\widehat{t}}+\left(  \frac{v^{2}}{1-v^{2}}\right)
(\mathbf{R}_{\widehat{\theta}\widehat{t}\widehat{\theta}\widehat{t}%
}+\mathbf{R}_{\widehat{\theta}\widehat{R}\widehat{\theta}\widehat{R}})\\
\mathbf{R}_{\widehat{3}^{\prime}\widehat{0}^{\prime}\widehat{3}^{\prime
}\widehat{0}^{\prime}}  &  =\mathbf{R}_{\widehat{\varphi}\widehat{t}%
\widehat{\varphi}\widehat{t}}+\left(  \frac{v^{2}}{1-v^{2}}\right)
(\mathbf{R}_{\widehat{\varphi}\widehat{t}\widehat{\varphi}\widehat{t}%
}+\mathbf{R}_{\widehat{\varphi}\widehat{R}\widehat{\varphi}\widehat{R}%
})\nonumber
\end{align}
Since, by spherical symmetry, $\mathbf{R}_{\widehat{\theta}\widehat{t}%
\widehat{\theta}\widehat{t}}=\mathbf{R}_{\widehat{\varphi}\widehat{t}%
\widehat{\varphi}\widehat{t}}$ and $\mathbf{R}_{\widehat{\theta}\widehat
{R}\widehat{\theta}\widehat{R}}=\mathbf{R}_{\widehat{\varphi}\widehat
{R}\widehat{\varphi}\widehat{R}}$, we get, plugging in the relevant
expressions [1]
\begin{align}
\left\vert \mathbf{R}_{\widehat{2}^{\prime}\widehat{0}^{\prime}\widehat
{2}^{\prime}\widehat{0}^{\prime}}\right\vert  &  =\left\vert \mathbf{R}%
_{\widehat{3}^{\prime}\widehat{0}^{\prime}\widehat{3}^{\prime}\widehat
{0}^{\prime}}\right\vert =\left\vert \frac{1}{2R^{2}(1-v^{2})}\left[
v^{2}\left(  b^{\prime}-\frac{b}{R}\right)  +2(R-b)\Phi^{\prime}\right]
\right\vert \nonumber\\
&  =\left(  \frac{2Br^{3}Z^{2}}{1-v^{2}}\right)  \times
\end{align}%
\[
\frac{\lbrack2Br(C+1)+\lambda(r^{2}+B^{2})(Cv^{2}+v^{2}-1)-2Br\lambda^{2}%
v^{2}]}{\lambda^{2}(r^{2}-B^{2})^{4}}.
\]
Assuming the lateral tidal force to be of the same order of magnitude as that
of the maximum radial tidal force, we get
\begin{equation}
\left\vert \mathbf{R}_{\widehat{2}^{\prime}\widehat{0}^{\prime}\widehat
{2}^{\prime}\widehat{0}^{\prime}}\right\vert =\left\vert \mathbf{R}%
_{\widehat{3}^{\prime}\widehat{0}^{\prime}\widehat{3}^{\prime}\widehat
{0}^{\prime}}\right\vert \lesssim10^{-2}cm^{-2},
\end{equation}
which leads to a particle velocity $v\sim0.66c$. This means that a freely
falling test particle accelerates from near zero velocity at one mouth to as
high as $0.66c$ at the radius $r=2.313$ and then decelerates to velocity
$0.18c$ at the throat $r_{0}^{+}=1.958$ before emerging into the other mouth.
Of course, the above assumption is not mandatory; the velocities will be less
and less as we go on decreasing the value on the right hand side of (31).
Overall, our conclusion is that the wormhole is traversible only
\textquotedblleft in principle\textquotedblright\ since the tidal forces are
finite and decay to zero at infinity, but not traversible in practice by humans.

\textbf{V. Total gravitational energy}

The total gravitational energy in localized sources having static spherical
symmetry and satisfying energy conditions is negative (attractive gravity). A
natural query is how the gravitational energy behaves under circumstances
where energy conditions are violated. To answer this, the known expression for
the gravitational energy has to be suitably adapted to account for situations
like the ones occurring in wormhole spacetimes.

All wormhole solutions require exotic material for their construction.
However, to our knowledge the gravitational energy content in the interior of
exotic matter distribution has not yet been studied. An intuitive approach in
this direction is provided by the formulation of gravitational energy by
Lynden-Bell, Katz and Bi\v{c}\'{a}k [30]. The energy formulation there is
intended for isolating and calculating the total attractive gravitational
energy $E_{G}$ of stationary gravity fields. In our view, their formulation
did not require any compelling restriction on the energy conditions of the
source matter.

The total gravitational energy $E_{G}$ appropriate for ordinary matter in a
normal star is given in [30] as%
\begin{equation}
E_{G}=Mc^{2}-E_{M}=\frac{1}{2}\int_{0}^{r}[1-(g_{rr})^{\frac{1}{2}}]\rho
r^{2}dr
\end{equation}
where the total mass-energy within the standard coordinate radius $r$ is
provided by Einstein's equations as%

\begin{equation}
Mc^{2}=\frac{1}{2}\int_{0}^{r}\rho r^{2}dr
\end{equation}
and the sum of other forms of energy like rest energy, kinetic energy,
internal energy etc is defined by
\begin{equation}
E_{M}=\frac{1}{2}\int_{0}^{r}(g_{rr})^{\frac{1}{2}}\rho r^{2}dr.
\end{equation}
The factor $\frac{1}{2}$ comes from $\frac{4\pi}{8\pi}$. Note that $E_{G}<0$
for ordinary matter configuration that has attractive gravity.

We shall adapt $E_{G}$ to wormhole geometry and distinguish it by
$\widetilde{E}_{G}$. By construction, the wormhole geometry has a hole instead
of a center and so we shall change the lower limit of integration in Eq.(32)
to the minimum allowed radius or throat $R_{0}$ defined by $b(R_{0})=R_{0}$.
The radius $R$ has the significance that it is the embedding space radial
coordinate; it decreases from $+\infty$ to $R=R_{0}$ in the lower side and
again increases to $+\infty$ in the upper side. This requires us to change the
integrals (33) and (34) to
\begin{align}
Mc^{2}  &  =\frac{1}{2}\int_{R_{0}}^{R}\rho R^{2}dR+\frac{R_{0}}{2}\\
E_{M}  &  =\frac{1}{2}\int_{R_{0}}^{R}(g_{RR})^{\frac{1}{2}}R^{2}dR,
\end{align}
the entire spacetime geometry being assumed to be free of singularities. The
constant $\frac{R_{0}}{2}$ in Eq.(35) comes from the integration of Einstein's
equation $\frac{\partial M}{\partial R}=\frac{1}{2}\rho R^{2}$ and we shall
choose it so as to offset the inner boundary term $\frac{b(R_{0})}{2}$ coming
from the integration. When $\rho=0$, we should fix $R_{0}=0$ in order to
recover $M=0$. In geometries with a regular center, one has $R_{0}=0$, the
above then reproduces Eqs.(33) and (34) respectively.

For wormholes, however, $R_{0}\neq0$. The difference between the above
integrals, viz.,
\begin{equation}
\widetilde{E}_{G}=Mc^{2}-E_{M}=\frac{1}{2}\int_{R_{0}}^{R}[1-(g_{RR}%
)^{\frac{1}{2}}]\rho R^{2}dR+\frac{R_{0}}{2}%
\end{equation}
is what we define here as the total gravitational energy of wormholes within
the region of integration [31]. Clearly, it is a straightforward adaptation of
Eq.(32) to wormhole geometry. However, one immediately notices that due to the
presence of the nonzero last term, the sign of $\rho$ does not necessarily
determine the sign of $\widetilde{E}_{G}$, as would be the case otherwise.

Looking at the MTY form (9), the Eq.(37) is rephrased on one side of the
wormhole as%
\begin{equation}
\widetilde{E}_{G}^{+}=\frac{1}{2}\int_{r_{0}^{+}}^{r_{1}}[1-(g_{RR})^{\frac
{1}{2}}]\rho R^{2}\frac{dR}{dr}dr+\frac{R_{0}^{+}}{2}%
\end{equation}
and on the other side as%
\begin{equation}
\widetilde{E}_{G}^{-}=-\frac{1}{2}\int_{r_{0}^{+}}^{r_{1}}[1-(g_{RR}%
)^{\frac{1}{2}}]\rho R^{2}\frac{dR}{dr}dr-\frac{R_{0}^{+}}{2}%
\end{equation}
where $r_{1}$ is an arbitrarily chosen fixed radius, $R_{0}^{+}$ is the throat
radius in $R-$coordinate to be computed from Eq.(10) and
\begin{equation}
g_{RR}=\left[  1-\frac{b(R)}{R}\right]  ^{-1}.
\end{equation}
Signature protection in the metric (9) requires that $1-(g_{RR})^{\frac{1}{2}%
}<0$, which, together with the information that $\rho<0$ should yield a
positive value of the integral in (38). All the factors in the integrand are
now known, but it turns out that an analytic integration is not possible. We
therefore do the numerical integration with $B=1$, $r=r_{0}^{+}$, $r_{1}=20$
(say), and using the Eqs.(10), (38)-(40), we find for typical values of
$\omega$ from Sec.III that%
\begin{align}
\omega &  =-5\text{, }r_{0}^{+}=1.96\text{, }R_{0}^{+}=5.82\text{, }%
\widetilde{E}_{G}^{+}=3.74=-\widetilde{E}_{G}^{-}\\
\omega &  =-1.4\text{, }r_{0}^{+}=2.92\text{, }R_{0}^{+}=8.26\text{,
}\widetilde{E}_{G}^{+}=5.50=-\widetilde{E}_{G}^{-},
\end{align}
which suggest that the total gravitational energy in the spherical shell is
positive indicating repulsive gravity. Likewise, $\widetilde{E}_{G}^{-}<0$ for
either value of $\omega$, implying attractive gravity.

\textbf{VI. Conclusions and remarks}

Static spherically symmetric wormholes in the Brans-Dicke theory have been
reasonably well discussed in the literature, but a thorough analysis of many
features is still lacking. In the present work, our aim was to fill that gap
obtaining several key results.

Our analysis fundamentally supports the basic constraint on the BD constants,
viz., $(C+1)^{2}>\lambda^{2}$, obtained earlier by Nandi \textit{et al}
[13,14], re-discussed and refined by Bloomfield [27,28]. The discussion in
Sec.III shows that much depends on the \textit{sign} of $\lambda$. We have
found a new range on the coupling parameter, viz., $-\frac{3}{2}<\omega
<-\frac{4}{3}$ for negative $\lambda$, and $\omega<-2$ for positive $\lambda$.
The latter range was obtained earlier by Agnese and La Camera [12] using a
different gauge. One might ask what would happen if we allow a thorough
mix-up, that is, if we interchange signs of $\lambda$ keeping the ranges for
$\omega$ uninterchanged, or even getting out of the suggested ranges? We will
then see that the throat radii become either negative or complex or fall below
the singular radius $B$. \ All these have to be disregarded as being
unphysical. Next, we have shown that both WEC\ and NEC are violated for the
ranges of $\omega$ specified above.

The next result is about traversibility. A remarkable aspect, heretofore
unnoticed, of the Brans class I solution is that it is the wormhole analogue
of the Horowitz-Ross naked black hole:\ An infalling observer meets the
maximum radial tidal force not at the throat but above it. It is then
quantitatively shown that Brans-Dicke wormhole can not be humanly traversible
because of the occurrence of tidal force $10^{18}$ times stronger than that on
Earth's surface. However, it is traversible \textquotedblleft in
principle\textquotedblright:\ An inanimate test particle can traverse through
it with a velocity profile as derived for the case $\omega=-5$ in Sec.IV.

We have shown in Sec.V that the gravitational energy content in an arbitrary
spherical shell around the wormhole throat is positive implying repulsive
gravity. The other side has attractive gravity. Hence the wormhole is like a
Janus faced object that sucks in matter at one mouth and spews out at the
other. All the above results might have important implications in astrophysics.

Finally, we wish to make some remarks clarifying certain issues:

(1) The solutions (5-8) describe a black hole, asymptotically flat only if
$\omega<-\frac{3}{2}$, and with a horizon of infinite surface area giving rise
to the so called cold black holes [32]. Hence, the black hole solutions in the
BD theory necessarily have a throat. But it is also possible to have pure
wormhole solutions. Campanelli \& Lousto [33] and Bronnikov [32] investigated
the general properties, including the structure and stability, of black hole
and wormhole solutions in the BD theory. We have argued above that wormhole
solutions arise even when the condition $\omega<-\frac{3}{2}$ is violated
[Sec.III, case (ii)]. It would be interesting to examine how the present work
relates to those just mentioned.

(2) We recall that the BD theory is Machian such that the scalar field
$\varphi$ is more like the zeroth component of gravitational potential than
any material field. In addition, it couples nonminimally in the
Hilbert-Einstein action. On the other hand, in the conformally rescaled
Einstein frame, the redefined scalar couples minimally and provides
unambiguous source stresses. So the procedure should be to first derive the
metric and corresponding stresses in the Einstein frame, then via certain
operations get to those in the Jordan frame. For instance, the Bronnikov-Ellis
solution [19] in the minimally coupled theory can be easily transferred into
Brans class I solution by redefining the constants together with inverse Dicke
transformations, as already shown in Ref.[22]. Under these operations, the
Einstein frame stress components readily convert into those [e.g., Eqs.(19)
and (24)] in the BD Jordan frame. In this sense, one talks about the stress
components of the BD scalar field, just as was done in Ref.[12]. In fact, the
definition of energy density and pressure is frame dependent: these quantities
have different behaviours in the so-called Einstein's frame, with gravity
minimally coupled to the scalar field, or in the \ Jordan's frame, with
gravity non-minimally coupled to the scalar field. Black hole solutions can
occur only when $\omega<-3/2$ and this implies a negative energy in all
space-time when the solution is transposed to the Einstein's frame. In this
frame the energy density of the scalar field is positive when $\omega>-3/2.$
The behaviour is different when the Jordan's frame is used. This may explain
why negative energy is obtained for the case $\omega=-1.4$, as shown in Fig.4 [34].

(3) A word of caution: During the discussion in Sec.III, we had taken the
limit $\omega\rightarrow\pm\infty$ to arrive at the Schwarzschild solution,
which is correct. But there is a prevailing belief that general relativity is
always recovered from BD theory in limit $\omega\rightarrow\pm\infty$. This is
now known not to be correct, see Refs.[35-38].

\bigskip

\textbf{Figure captions:}

Fig.1. The simultaneous inequalities $r_{0}^{+}>B$ and $f=(C+1)^{2}%
-\lambda^{2}>0$ determine the range $\omega<-2$, when $\lambda$ is positive.
In all the figure we have set units in which $B=1$. The functions $r_{0}^{+}$
and $f$ are expressed in terms of $\omega$ using Eqs.(13), (22) and (23).

Fig.2. WEC and NEC violation for a range of $r\in\lbrack r_{0}^{+}$,$20]$ and
typical value, $\omega=-5$ [case (i)]. See Eqs.(19) and (25).

Fig.3. The simultaneous inequalities $r_{0}^{+}>B$ and $f=(C+1)^{2}%
-\lambda^{2}>0$ determine the range $-\frac{3}{2}<\omega<-\frac{4}{3}$, when
$\lambda$ is negative.

Fig.4. WEC and NEC violation for a range of $r\in\lbrack r_{0}^{+}$,$20]$ and
typical value, $\omega=-1.4$ [case (ii)]. See Eqs.(19) and (25).

Fig.5. Plot of $g(r)$ vs $r$ for a range of $r\in\lbrack r_{0}^{+}$,$10]$ and
typical value, $\omega=-5$ [case (i)]. See Eq.(28).

Fig.6. Plot of $g(r)$ vs $r$ for a range of $r\in\lbrack r_{0}^{+}$,$10]$ and
typical value, $\omega=-1.4$ [case (ii)] See Eq.(28).

\bigskip

\textbf{Acknowledgments:}

The authors are grateful to Ms. Sonali Sarkar of NBU for her assistance with
figures. We thank two anonymous referees whose insightful comments have led to
many improvements on the article.

\bigskip

\textbf{References}

[1] M. S. Morris and K. S. Thorne, Am. J. Phys. \textbf{56}, 395 (1988); M. S.
Morris, K. S. Thorne, and U. Yurtsever, Phys. Rev. Lett. \textbf{61}, 1446 (1988).

[2] A. Einstein and N. Rosen, Phys. Rev. \textbf{48}, 73 (1935).

[3] J.A. Wheeler, \textit{Geometrodynamics} (Academic Press, New York, 1962).

[4] J.G. Cramer, R.L. Forward, M.S. Morris, M. Visser, G. Benford, and G.A.
Landis, Phys. Rev. D \textbf{51}, 3117 (1995).

[5] V. Bozza, Phys. Rev. D \textbf{66}, 103001 (2002).

[6] K.S. Virbhadra and G.F.R. Ellis, Phys. Rev. D \textbf{62}, 084003 (2002).

[7] A. Bhadra, Phys. Rev. D \textbf{67},103009 (2003).

[8] K.K. Nandi, Y.Z. Zhang, and A.V. Zakharov, Phys. Rev. D \textbf{74},
024020 (2006).

[9] D. Hochberg, A. Popov, and S. V. Sushkov, Phys. Rev. Lett. \textbf{78},
2050 (1997); Nail R. Khusnutdinov, Phys. Rev. D \textbf{67}, 124020 (2003).

[10] N.S. Kardashev, I.D. Novikov, and A.A. Shatskiy, Int. J. Mod. Phys. D
\textbf{16}, 909 (2007).

[11] M. Visser, \textit{Lorentzian Wormholes-From Einstein To Hawking} (AIP,
New York, 1995).

[12] A.G. Agnese and M. La Camera, Phys. Rev. D \textbf{51,} 2011 (1995).

[13] K.K. Nandi, A. Islam, and J. Evans, Phys. Rev. D \textbf{55}, 2497 (1997).

[14] K.K. Nandi, B. Bhattacharjee, S.M.K. Alam, and J. Evans, Phys. Rev. D
\textbf{57}, 823 (1998).

[15] L.A. Anchordoqui, S. P. Bergliaffa, and D.F. Torres, Phys. Rev. D
\textbf{55,} 526 (1997).

[16] A. Bhadra and K. Sarkar, Mod. Phys. Lett. A \textbf{20, }1831 (2005).

[17] F.S. N. Lobo, Phys. Rev. D \textbf{71}, 084011 (2005); Phys. Rev. D
\textbf{73}, 064028 (2006).

[18] S. V. Sushkov, Phys.Rev. D \textbf{71}, 043520 (2005).

[19] K.A. Bronnikov, Acta Phys. Polon. B\textbf{\ 4}, 251 (1973); H.G. Ellis,
J. Math. Phys. \textbf{14}, 104 (1973).

[20] E.F. Eiroa and C. Simeone, Phys. Rev. D \textbf{76}, 024021 (2007).

[21] F. Rahaman, M. Kalam, and S. Chakraborty, Gen. Rel. Grav. \textbf{38},
1687 (2006).

[22] K.K. Nandi, I. Nigmatzyanov, R. Izmailov, and N.G. Migranov, Class.
Quant. Grav. \textbf{25}, 165020 (2008).

[23] J.P.S. Lemos, F.S.N. Lobo, and S.Q. de Oliveira, Phys. Rev. D \textbf{68}
064004 (2003).

[24] C. H. Brans, Phys. Rev. \textbf{125}, 2194 (1962).

[25] J.L. Friedman, K. Schleich, and D.M. Witt, Phys. Rev. Lett. \textbf{71},
1486 (1993).

[26] D. Hochberg and M. Visser, Phys. Rev. Lett. \textbf{81}, 746 (1998).

[27] P.E. Bloomfield, Phys. Rev. D \textbf{59}, 088501 (1999).

[28] K.K. Nandi, Phys. Rev. D \textbf{59}, 088502 (1999).

[29] G.T. Horowitz and S.F. Ross, Phys. Rev. D \textbf{56}, 2180 (1997);

[30] D. Lynden-Bell, J. Katz, and J. Bi\v{c}\'{a}k, Phys. Rev. D \textbf{75},
024040 (2007).

[31] K.K. Nandi, Y.Z. Zhang, R.G. Cai, and A. Panchenko, Phys. Rev. D
\textbf{79}, 024011 (2009).

[32] K.A. Bronnikov, G. Cl\'{e}ment, C.P. Constantinidis, and J.C. Fabris,
Phys. Lett. A \textbf{243}, 121 (1998).

[33] M. Campanelli and C.O. Lousto, Int. J. Mod. Phys. D \textbf{2}, 451 (1993).

[34] We thank an anonymous referee for the explanation.

[35] V. Faraoni, \textit{Cosmology in Scalar-Tensor Gravity}, Ch.1, p.13
(Kluwer Academic Publishers, 2004) and references therein.

[36] C. Romero and A. Barros, Phys. Lett. A \textbf{173}, 243 (1993).

[37] N. Banerjee and S. Sen, Phys. Rev. D \textbf{56}, 1334 (1997).

[38] A. Bhadra and K.K. Nandi, Phys. Rev. D \textbf{64}, 087501 (2001).

\bigskip\newpage

\begin{center}

\end{center}

\end{document}